\documentstyle[multicol,aps,prl,epsf,psfig]{revtex}
\begin{document}
\title{ Phase Transition in
the Takayasu Model with Desorption}
\author{Satya N. Majumdar$^{1}$, Supriya Krishnamurthy$^{2}$ and
 Mustansir Barma$^{1}$ } 
\address{ 
{$1$ Tata Institute of Fundamental Research, Homi
 Bhabha Road,  Mumbai-400005, India }\\
{$2$ Department of Theoretical Physics, University of Oxford,
1 Keble Road, Oxford, OX1 3NP, UK} \\
}
\maketitle

\begin{abstract}
We study a lattice  
model where particles carrying different masses diffuse,
coalesce upon contact, and also unit masses adsorb to a site with rate
$q$ or desorb from a site with nonzero mass with rate $p$.
In the limit $p=0$ (without desorption), our model reduces to the
well studied Takayasu model where the steady-state single site
mass distribution has a power law tail $P(m)\sim m^{-\tau}$ for
large mass.
We show that varying the desorption rate $p$ induces
a nonequilibrium phase transition in all dimensions.   
For fixed $q$, there is a critical $p_c(q)$ such that if $p<p_c(q)$,
the steady state mass distribution, $P(m)\sim m^{-\tau}$ for large $m$
as in the Takayasu case. For $p=p_c(q)$, we find $P(m)\sim m^{-\tau_c}$
where $\tau_c$ is a new exponent, while for $p>p_c(q)$, $P(m)\sim
\exp(-m/m^*)$ for large $m$. The model is studied analytically
within a mean field theory and numerically in one dimension.
\end{abstract}
\vskip 1cm


\section{Introduction}
Many systems in nature, ranging from reaction-diffusion systems
to fluctuating interfaces, exhibit nonequilibrium steady states
with a wide variety of phases. Of particular interest are the
self-organized critical systems where different physical quantities
have power law distributions in the steady state over a wide
region of the parameter space\cite{BTW}. Self-organized criticality has
been studied
in a variety of model systems ranging from sandpiles to earthquakes.
A particularly simple lattice model due to Takayasu, where masses diffuse,
aggregate upon contact and adsorb unit masses from outside at a constant
rate, was shown to exhibit self-organized criticality\cite{Takayasu}:
the steady-state mass distribution has
a nontrivial power law decay for large mass in all
dimensions\cite{Takayasu}. This
model initially generated a lot of attention as it was a
simple exactly solvable model of self organized criticality with close
connections\cite{dd} to 
other solvable models such as the Scheidegger river
model\cite{scheidegger}, the voter
model\cite{liggett} and the directed abelian sandpile model\cite{ddrr}.
Recently there has been a renewed interest in this model as simple
variants of the Takayasu model have been found useful in modelling
the dynamics of a variety of systems including
force fluctuations in granular systems such
as bead packs\cite{SC}, river networks\cite{AM},
voting systems\cite{Melzak,FF}, wealth distributions\cite{IKR},
inelastic collisions in
granular gases\cite{Haff}, the generalized Hammersley
process\cite{aldous}, particle systems in one dimension\cite{krug}
and various generalized mass transport models\cite{RM}.

In the Takayasu model, each site of a lattice has a nonnegative mass
variable. Starting from an initial random distribution of masses, each
mass hops to a nearest neighbour site (chosen at random) and aggregates
with the mass there with rate $1$. In addition, a unit mass is adsorbed
at every site with rate $q$. While the first move tends to create
big masses via diffusion and aggregation, the second move replenishes
the lower end of the mass spectrum. At large time $t$, the mass
distribution at any site has the scaling behaviour, $P(m,t)\sim
m^{-\tau}f(m/t^{\delta})$ with $\delta=1/(2-\tau)$\cite{T4,MS}. The
interesting point is that even though the average mass per site increases
linearly with time, $\langle m \rangle \sim t$, the mass distribution
$P(m,t)$ approaches a time-independent power law distribution $P(m)\sim
m^{-\tau}$ for $t\to \infty$ (since $f(0)\sim O(1)$) for any nonzero
adsorption rate $q$. The exponent $\tau$ is independent of $q$ and is
known exactly\cite{Takayasu},
$\tau=4/3$ in one dimension and $\tau=3/2$ within mean field theory.   
 
The steady-state mass distribution in the Takayasu model has 
the same power law decay for any nonzero adsorption rate $q$
and does not undergo any phase transition. In this paper we show 
that if we introduce an additional process of desorption of unit masses
with rate $p$ in the Takayasu model (we call this the In-out model), a
rich steady-state phase diagram
emerges in the $p-q$ plane. In particular
we show that the system
undergoes a nonequilibrium phase transition across a phase boundary
$p_c(q)$. Nonequilibrium phase transitions between
steady states have been studied extensively in recent years in a variety
of systems. Examples include, amongst others, active-absorbing phase
transitions in reaction
diffusion systems\cite{RD}, roughening transitions in fluctuating
interfaces\cite{JK}, phase transitions in driven diffusive lattice
gas models\cite{SZ},
wetting transitions in solid-on-solid models\cite{hinrichsen},
boundary driven transitions in one dimensional asymmetric exclusion
processes\cite{DEHP} and 
Bose-Einstein like condensation in
models of aggregation and fragmentation\cite{MKB,KR}. However we show
below that the mechanism of the phase transition and the
associated critical properties in the In-out
model are very different from those of other models mentioned
above.

There are quite a few physical systems where our In-out model may find
applications. In nature there exist a variety of systems ranging from 
colloids\cite{White} to polymer gels\cite{gel} where the
basic
constituents of the system diffuse and coalesce upon contact.
For example, in a polymer gel the basic constituents are polymers
of different sizes which diffuse in a solution and when an $m$-mer comes
in contact with an $n$-mer, they aggregate to form an
$(m+n)$-mer\cite{gel}.
Similarly during the growth of a thin film on an amorphous substrate (such
as Bismuth on Carbon), clusters or islands of atoms can diffuse as a
whole and when two of them come closer they
coalesce\cite{thinfilms}. A zeroth order approach to model 
the dynamics of these systems would be to replace each cluster by a point
particle (ignoring its shape) carrying a positive mass which
indicates its size or number of atoms. When two particles coalesce
their masses add up. In addition many of these systems are
{\it open} in the sense that they can exchange basic units with the
adjoining environment. For example, during the growth of a film on a
substrate, single adatoms may adsorb on the substrate from the outside
vapour or desorb into the vapour from the substrate. 
We attempt to incorporate
these processes on a lattice in the In-out model and show that
even this simple model has a very rich steady-state phase diagram.
We had introduced this model in 
an earlier publication\cite{MKB} and some results were briefly
mentioned. In this paper we present a detailed analysis of the model.

The paper is organized as follows. In section II, we define the In-out
model precisely and summarize the different phases and the transitions
between them. In section III, we solve the model analytically within mean
field theory.
In section IV, we present the numerical results in one dimension
and discuss a scaling theory which provides scaling relations between
different critical exponents. We conclude in section V with a summary and
a discussion of open questions.

\section{The In-out Model}
For simplicity we define the In-out model on a one-dimensional lattice
with periodic boundary conditions; generalizations to higher dimensions
are straightforward. Each site of a lattice has a nonnegative
mass variable $m_i\ge 0$. Initially each $m_i$ is chosen independently
from any well defined distribution. The dynamics proceeds as follows.
A site $i$ is chosen at random and then one of the following 
events can occur: 
\begin{enumerate}
\item Adsorption:
With probability $q/(p+q+1)$, a single particle is adsorbed at site
$i$; thus $m_i \rightarrow m_{i}+1$.
\item Desorption 
With probability $p/(p+q+1)$, a single particle detaches from and 
leaves site
$i$; thus $m_i \rightarrow m_{i}-1 $ provided  $m_i \ge 1$.
\item Diffusion and Aggregation:
With probability $1/(p+q+1)$, the mass $m_i$ at site $i$ moves to a
nearest neighbour site [either $(i-1)$ or $(i+1)$] chosen at random.
If it moves to a site which already has some 
particles, then the total mass just adds up; thus $m_i \rightarrow 0$ and
$m_{i\pm1}\rightarrow m_{i\pm1}+m_i$.
\end{enumerate}
If the site chosen is empty, only adsorption can occur with 
probability $q/(p+q+1)$. 

The In-out model has only two parameters $p$ and $q$. The question
we would like to address is: For given $p$ and $q$, what is the single
site mass distribution $P(m)$ in the steady state?
Note that in the limit $p=0$ (i.e., without the desorption process)
our model reduces to the Takayasu model mentioned in the introduction.

While the Takayasu model (zero
desorption, $p=0$) does not have a phase transition in the steady state,
we find that introducing a nonzero desorption rate $p$ induces a rich
steady state behaviour in the $p-q$ plane. In fact we find that there is
a critical line $p_c(q)$ in the $p-q$ plane. For fixed $q$, if we
increase
$p$ from $0$, we find that for all $p<p_c(q)$, the steady state mass
distribution has the same large $m$ behaviour as in the Takayasu
case, i.e., $P(m)\sim m^{-\tau}$ where the exponent $\tau$ is the Takayasu
exponent and is independent of $q$. Thus the Takayasu phase is 
stable upto $p_c$. For $p=p_c(q)$, we
find the steady state mass distribution still decays algebraically for
large $m$, $P(m)\sim m^{-\tau_c}$ but with a new critical exponent
$\tau_c$ which is bigger than the Takayasu exponent $\tau$. For
$p>p_c(q)$, we find that $P(m)\sim \exp (-m/m^{*})$ for large $m$ where
$m^{*}$ is a characteristic mass that diverges if one approaches $p_c(q)$
from the $p>p_c(q)$ side. The critical exponent $\tau_c$ is the same
everywhere on the critical line $p_c(q)$. This phase transition occurs
in all spatial dimensions including $d=1$. 

It is easy to write down an exact
evolution equation for the mean mass $\langle m \rangle(t)$ per site.
Since the diffusion and aggregation move does not change the total mass,
the only contributions to the time evolution of $\langle m\rangle$ come
from the adsorption and desorption processes. It is then evident that, 
\begin{equation}
\frac{d\langle m\rangle}{dt}=q-ps(t)
\label{eq:mass}
\end{equation}
where $ s(t)$ is the probability that a site is occupied by a
nonzero mass. The first term on the right hand side of the above
equation clearly indicates the increase in
mass per site due to the adsorption of unit mass. The
second term quantifies the loss in mass per site due to the
desorption of unit mass taking into acount the fact that the
desorption can take place from a site only if the site is occupied 
by a nonzero mass. Let us fix $p$ and vary $q$. As long as $q<q_c(p)$,
it turns out that in the long time limit $t\to \infty$, the two terms
on the right hand side of the above equation cancel each other and
the occupation density reaches the asymptotic time independent value,
$s=q/p$. This indicates that the average mass per site, $\langle m\rangle$
becomes a constant in the long time limit. In fact, we show below that in
this phase, the steady state mass distribution $P(m)\sim \exp(-m/m^{*})$
for large $m$ with a finite first moment $\langle m\rangle$. We call this
phase the ``Exponential " phase. However
if $q>q_c(p)$, the occupation density reaches a steady state value $s$
such that $s<q/p$. As a result in the long time limit, the
second term on the right hand side of Eq. (1) fails to cancel the first 
term and the mean mass per site $\langle m \rangle (t)$ increases
linearly with time, $\langle m\rangle \sim (q-ps)t$. However, as we show
below, even though the mean mass diverges in this phase as $t\to
\infty$, the mass distribution reaches a steady state, $P(m)\sim
m^{-\tau}$ for large $m$ where $\tau$ is the Takayasu exponent (which is
always less than $2$ so that the mean mass diverges). Hence
we call this entire phase the ``Takayasu" phase.

\section{Mean Field Theory}

We first analyze the model exactly within the mean field approximation,
ignoring correlations in the occupancy of adjacent sites. In that case we
can directly write down equations for $ P(m,t)$, the probability that any
site has a mass $m$ at time $t$.
\begin{eqnarray}
\frac{dP(m,t)} {dt}= &-&(1+p+q+s)P(m,t) 
+ p P(m+1,t)  \nonumber \\
&+& q P(m-1,t)+ P*P ;\;\;\; m \geq 1~~ \label{eq:mft1}\\
\frac{dP (0,t)} {dt} = &-& (q+s) P(0,t) + p P(1,t) + 
s(t) \label{eq:mft2}. 
\end{eqnarray} 
Here $P*P=\sum_{m^{\prime}=1}^{m}P(m^{\prime},t)P(m-m^{\prime},t)$ is a
convolution term that describes the coalescence of two masses and
$s(t)=\sum_{m=1}P(m,t)$ denotes the probability that a site is occupied 
by a nonzero mass. 

The above equations enumerate the possible ways in  which the  mass 
at a site might change. The first term in Eq. (\ref{eq:mft1}) is
the ``loss'' term that accounts for the probability that a
mass $m$ might move as a whole or desorb or adsorb a unit mass,
or a mass from the neighbouring site might move to the site in 
consideration. In this last case,
the probability of occupation of the neighbouring site, 
$s(t)$ multiplies $P(m,t)$ within the mean-field
approximation where one neglects the spatial correlations in the
occupation probabilities of neighbouring sites. The remaining three terms
in Eq. (\ref{eq:mft1}) are the ``gain'' terms enumerating the number of ways
that a site with mass $m^{\prime} \neq m$ can gain or lose mass to make 
the final mass $m$. The second equation Eq. (\ref{eq:mft2}) is a similar
enumeration of the possibilities for loss and gain of empty sites. 

To solve the equations, we compute the generating function, 
$Q(z,t) = \sum_{m=1}^{\infty} P(m,t)z^{m}$ from Eq. (\ref{eq:mft1}) and set 
$ \partial Q / \partial t =0$ in the steady state. We also need to use
Eq. (\ref {eq:mft2}) to write $P(1,t)$ in terms of $s(t)$. This gives
us a quadratic equation for $Q$ in the steady state.
Choosing the root that corresponds to $Q(z=0) = 0$, we find
\begin{equation}
2zQ(z) = p(z-1)
+ qz(1-z)+2sz-\sqrt {(z-1)\Delta(z)}.  \label{eq:qsol}
\label{eq:detm}
\end{equation}
where 
\begin{eqnarray}
\Delta(z)=&&p^2(z-1)+q^2z^2(z-1)-2pqz(z-1)\nonumber\\
         &-&4qz(z-sp/q).
\label{eq:delta}
\end{eqnarray}
Note that the occupation density $s$ in the above expression of $Q(z)$
is yet to be determined. The steady state mass distribution $P(m)$ 
can be formally obtained from $Q(z)$ in Eq. (\ref{eq:qsol})
by evaluating the Cauchy integral, 
\begin{equation}
P(m) = {1\over {2\pi i}}\int_{C_o} \frac {Q(z)} {z^{ m+1}} dz
\label{eq:contour}
\end{equation}
over a contour $C_o$ encircling the origin in the complex plane. This
expression for $P(m)$
however will contain the yet to be determined unknown quantity $s$.
In fact, determining $s$ is the most nontrivial
part of the mean field calculation as we show below.  

In order to extract the large-$m$ behaviour of $P(m)$ from
Eq. (\ref{eq:contour}), one needs to deform the contour $C_o$
so that it goes around the branch cut singularities of the function
$Q(z)$. From Eq. (\ref{eq:detm}), it is evident that such singularities 
occur at $z=1$ and also at the roots of $\Delta(z)=0$
where $\Delta(z)$ is given by Eq. (\ref{eq:delta}). Since $\Delta(z)$
is a cubic polynomial in $z$, it has three roots $z_1$, $z_2$ and $z_3$,
each of which can be determined in terms of the unknown quantity $s$.

We now analyse the large $m$ behaviour of $P(m)$ in different regions of
the $p-q$ plane. Let us fix the value of $p$ and increase $q$ from $0$.
A similar analysis can be carried out for fixed $q$ as a function of
$p$.
As we increase $q$ from $0$, we encounter the following three regimes,
\hfill\break
 
(i) For small $q$ (with a fixed $p$), we first assume that the mean mass
$\langle m\rangle$ reaches a time-independent constant as $t\to \infty$.
This assumption will be justified {\it a posteriori}. Then from Eq. (1),
it follows that the occupation density also reaches a steady state
value, $s=q/p$. Substituting
this in the expression for $\Delta(z)$ in Eq. (\ref{eq:delta}), 
the three roots of $\Delta(z)=0$  
are $z_1=1$ and $z_{2,3}=(p+2 \mp 2\sqrt{p+1})/q$. 
Then from Eq. (\ref{eq:detm}), it follows that the only branch cut
singularities of $Q(z)$ are at $z_2$ and $z_3$ with $z_3>z_2>1$ for small
$q$. Therefore the branch cut at $z_2$ essentially controls the large-$m$
behaviour of $P(m)$ when the contour in Eq. (\ref{eq:contour}) is deformed
and by analysing the integral around this cut we find that for large $m$,
\begin{equation}
P(m) \sim exp(-m/m^*)/m^{3/2} 
\end{equation}
with $m^*=1/ln z_2$. Since
$P(m)$ decays exponentially in this phase, $\langle m\rangle$ is also
finite and nonzero thus justifying the assumption made
in the begining. In this phase the unknown function $s$ is therefore
exactly given as $s=q/p$. Note however that this
analysis is valid as long as $z_2>1$ and the characteristic mass $m^{*}$
diverges as $z_2$ approaches $1$ from above. \hfill\break

(ii) As the value of $q$ is increased (for fixed $p$) 
the roots $z_2$ and $z_3$ decrease, until at a critical
value $q_c(p)$, the value of $z_2$ just reaches unity.  The double root
($z_1$ and $z_2$) at $z=1$ of $\Delta(z)$ then leads to a branch cut
singularity of
order $3/2$ in $Q(z)$ in Eq. (\ref{eq:detm}), which in turn implies 
\begin{equation}
P(m) \sim m^{-5/2}.
\end{equation}
This power law decay characterizes the critical point and the condition
$z_2=1$ determines the locus of the critical line in the $p-q$ plane,
\begin{equation}
q_c(p) =p+2-2{\sqrt {p+1}}.
\label{eq:phbou}
\end{equation}
The value of $s$ is given exactly by $s=q_c/p$.
\hfill\break

(iii) As $q$ is increased further ($q>q_c(p)$) for fixed $p$, the mean
mass per site $\langle m\rangle $ does not
reach a time-independent value in the steady state, but increases
indefinitely
with time.
Consequently we cannot use the relation $s=q/p$ anymore.  
However, $P(m)$ reaches a time-independent distribution. So the question
is what is the selection principle that determines the unknown function
$s$ in this regime?

Note that at $q=q_c(p)$, the two roots $z_1$ and $z_2$ of $\Delta(z)=0$
coincided, $z_1=z_2=1$ and $z_3>1$. As $q$ increases further, since
we do not know what $s$ is {\it a priori}, the exact locations of the
three roots of $\Delta(z)=0$ in the complex plane are also unknown. 
However since $\Delta(z)$ is a polynomial with real coefficients,
if $z$ is a root of $\Delta(z)=0$, so must be its complex conjugate $z^*$.
Thus as $q$ increases beyond $q_c(p)$, there are two possibilities.
The first possibility is that all the three roots of $\Delta(z)=0$ are
real and distinct. But in that case, as $q$ increases slightly beyond
$q_c(p)$, at least one of them must become less than $1$. This however
would lead to an exponential
growth of $P(m)$ for large $m$ and hence is ruled
out. The second and only possibility is that
one of the three roots must be real while the other
two are complex conjugates of each other, i.e.,     
$\Delta(z)=(z-z_c)(z-{z_c}^*)(z-z_3)$ where $z_3$ is real and
$z_c$ in general is complex with its real part less than $1$. However, if
the imaginary part of $z_c$
is nonzero, this again can be shown to lead to an exponential divergence
of $P(m)$ for large $m$. Therefore, we are led to the conclusion that
$z_c$ must be real and thus $\Delta(z)=0$ must have {\it double} roots at
$z_c$, i.e.,  
$\Delta(z)=(z-z_c)^2(z-z_3)$ with $z_c$ real. In summary we conclude that
for $q>q_c(p)$,
$z_3$ remains $>1$ and the two roots $z_1=z_2=z_{c}$ continues
to be coincident and real but the
common value $z_{c}$ decreases below $1$ as $q$ increases beyond
$q>q_c(p)$. This nontrivial `root
sticking' condition determines the unknown quantity $s$ for
$q>q_c(p)$. This condition of double roots can be easily implemented
by demanding the two conditions,
$\Delta (z_{c})=0$ and $\Delta^{\prime}(z_{c})=0$ where
$\Delta^{\prime}=d\Delta(z)/dz$. Also using the relation
$\Delta(z)=(z-z_{c})^2(z-z_3)$ in Eq. (\ref{eq:detm}), we find that the
lowest branch
cut singularity of $Q(z)$ is at $z=1$. This 
order $1/2$ singularity
then leads to the following asymptotic behaviour of $P(m)$,
\begin{equation}
P(m) \sim m^{-3/2} .
\end{equation}
Thus this entire phase, $q>q_c(p)$ is characterized 
by the same power-law decay of $P(m)$
as in the mean field Takayasu model which, as mentioned earlier,
corresponds to the zero-desorption ($p=0$) limit
of our model. 

As mentioned above, the `root-sticking' condition also
determines quite
non-trivially the occupation density $s$ for $q>q_c(p)$ for fixed $p$. 
To determine $s$ explicitly for $q>q_c(p)$ using this condition,
let us fix $p=1$ for
simplicity even though the calculation can be carried out for any
arbitrary $p$. From Eq. (\ref{eq:phbou}), we find
$q_c=3-2{\sqrt 2}$ for $p=1$. We first
substitute the
expression
for $\Delta(z)$ from Eq. (\ref{eq:delta}) in the `root-sticking'
conditions, $\Delta(z_{c})=0$ and $\Delta^{\prime}(z_{c})=0$. We
then eliminate $z_{c}$ from these equations and find
$s(q)$ for $q>3-2{\sqrt 2}$ as the only positive root of the cubic
equation
\begin{eqnarray}
16s^3&-&(q^2-12q+24)s^2-(q^3+5q^2+57q+15)s\nonumber\\
&+&(q^3+5q^2+39q-2)=0.
\label{eq:cubic}
\end{eqnarray} 
Thus we can determine the unknown quantity $s$ exactly everywhere
in the $p-q$ plane. In  Fig. 1, we plot the function $s(q)$ for fixed
$p=1$.  For $q\leq q_c=3-2{\sqrt 2}$ we have
$s(q)=q$ and for $q>q_c=3-2{\sqrt 2}$, $s(q)$ is
given by the real positive root of the
cubic equation in Eq. (\ref{eq:cubic}).

Note that for fixed $p$, if $q>q_c(p)$, the steady state value of $s(q)$
(as determined from the `root sticking' conditions) is less than $q/p$
and hence from Eq. (1), we find that the mean mass per site increases
linearly with time, $\langle m\rangle\approx vt$ for large $t$.
If one interprets the mass profile as the height of an interface
(see Section V) then for $q<q_c(p)$, the average ``height" of the
interface becomes a constant as $t\to \infty$, while for $q>q_c(p)$,
the average ``height" $\langle m\rangle$ increases linearly with
velocity $v$.  
The `velocity' $v$ defined more precisely as $v=\lim_{t\to
\infty}{{\langle
m\rangle}\over {t}}$ is $0$ for $q<q_c(p)$ and nonzero for $q>q_c(p)$.
For $q$ slightly bigger than $q_c(p)$, $v\sim [q-q_c(p)]^y$ where
$y$ is a critical exponent independent of $p$. For example for $p=1$,
we find from Eq. (\ref{eq:cubic}), $v\approx (q-q_c(1))^2/(6{\sqrt 2}-8)$
indicating that $y=2$ within mean field theory.

\section{Numerical results in one dimension and scaling theory}

Having completed the mean field calculations we now turn to one dimension.
While the Takayasu model ($p=0$) is exactly solvable in
$d=1$\cite{Takayasu}, the same technique unfortunately does not work for
$p>0$. Hence for nonzero $p$, we had to resort to numerical simulations 
in $d=1$. The qualitative predictions of mean field theory
namely the existence of
a power-law (Takayasu) phase ($P(m) \sim m^{-\tau_T}$)
and a phase with exponential mass
distribution, with a different critical behaviour at the transition ($P(m) \sim
m^{-\tau_c}$), are found to hold in 1-d as well. 
Figure 2 shows the results of numerical
simulations for the phase diagram along with the mean-field
prediction (Eq. \ref{eq:phbou}) and Figure 3 displays the numerical
data for the decay
of the mass distribution $P(m)$ in the two phases and
at the transition point.
The values obtained, $\tau=4/3$ (same as the exactly solvable $p=0$ case) 
and $\tau_c \simeq 1.833$,  
are quite different from 
their mean-field values $\tau=3/2$ and $\tau_c=5/2$, reflecting the
effects of correlations between 
masses at different sites.

If the phase boundary is crossed by increasing $q$ for fixed $p$, the
Takayasu phase is obtained for $q > q_c$. As a function of the
small deviation
$\tilde q \equiv q - q_c$ and large time $t$, the mass distribution $P(m,
\tilde
q, t)$ is expected to display a scaling form for large $m$,
\begin{equation}
P(m, \tilde q, t) \sim {1 \over m^{\tau_c}}
\ Y (m \tilde q^\phi, {m \over
t^\alpha})
\label{eq:scform}
\end{equation}
in terms of three unknown exponents $\phi$, $\alpha$, $\tau_c$ and
the two variable scaling function $Y$. All other exponents then can be 
related to these three exponents via scaling relations.
We give some examples below. \hfill\break

(a) Consider $\tilde q>0$ and $t\to \infty$ limit. Then $P(m,\tilde q)\sim
{1 \over m^{\tau_c}}Y(m \tilde q^\phi,0)$. But we know that for $\tilde
q>0$, in the steady state, $P(m, \tilde q)\sim m^{-\tau}$ where $\tau$ is
the known Takayasu exponent. This forces the scaling function $Y(x,0)\sim
x^{\gamma}$ for large $x$ such that $P(m,\tilde q)\sim {\tilde
q}^{\phi\gamma}/m^{\tau_c-\gamma}$, indicating  $\gamma=\tau_c-\tau$.
\hfill\break

(b) Consider again $\tilde q>0$ and finite but large $t$. The mean mass
per site, $\langle m\rangle =\int m P(m,\tilde q,t)dm \sim {\tilde q}^y t$
where $y$ is the `velocity' exponent. Using the scaling form of $P$, we
find, $y=\phi[1-\alpha(2-\tau_c)]/{\alpha}$.\hfill\break

(c) Next we consider the critical point, $\tilde q=0$. Using the scaling
form, we find that the mean mass, $\langle m\rangle \sim t^{\zeta}$ for
large $t$ where $\zeta=\alpha(2-\tau_c)$ provided $\tau_c<2$. If
$\tau_c>2$ (as in mean field theory), $\zeta=0$. Also, the root mean
square
mass fluctuations at the critical point, $\sigma=\sqrt {\langle (m-\langle
m\rangle)^2\rangle}\sim \sqrt{\langle m^2\rangle}\sim t^\beta$ for large
$t$ with $\beta=\alpha(3-\tau_c)/2$. Note that for large $t$, $\langle
m^2\rangle >>{\langle m\rangle}^2$ indicating that fluctuations grow
faster than the mean as time increases. \hfill\break

Within mean field theory, by analysing $P(m)$ explicitly for $\tilde q>0$,
we find $P(m,\tilde q)\sim {\tilde q}/m^{3/2}$ and also $\tau_c=5/2$. 
From (a) above, this 
immediately gives, $\gamma\phi=1$ and $\gamma=1$ indicating $\phi=1$.
Also, we had shown before that the velocity exponent $y=2$ exactly
within mean field theory. Using $y=2$, $\tau_c=5/2$ and $\phi=1$ in 
(b) of the previous paragraph, we get $\alpha=2/3$. Since $\tau_c=5/2>2$,
we note from (c) that $\zeta=0$. Also we find the fluctuation exponent
$\beta=1/6$ from the scaling relation in (c). Thus within mean field
theory, we find
\begin{equation}
P(m, \tilde q, t) \sim {1 \over m^{5/2}} \ Y (m \tilde q, {m \over
t^{2/3}}).
\label{eq:mfscform}
\end{equation}

We have determined the corresponding exponents in $d=1$ numerically.
The critical exponent $\tau_c\simeq 1.83$ has already been mentioned
(see Fig 3). In Fig 4, we plot the velocity $v$ as a function of $q$
for fixed $p=2.35$. The velocity is zero for $q\leq q_c\approx 1.0$
and increases as a power law, $v\sim (q-q_c)^y$ for small $\tilde
q=(q-q_c)$. We find $y\simeq 1.47$. Note that since $q_c$ is not known
exactly, this exponent is difficult to determine numerically and is
subject to large error bars. We also find that at the critical point
$q_c\approx 1$, the mean mass grows as, $\langle m\rangle \sim t^{\zeta}$
with $\zeta\simeq 0.12$. Note the difference from the mean field theory
where $\langle m\rangle$ does not grow with time at the critical point
($\zeta=0$). To measure the fluctuations at the critical point,
we performed finite-size studies of
the time-dependent `width' $W^2(t,L)={\sum_{i=1}^{L} (m_i-<m>)^2}/L$ 
at the critical point, where $L$ is the system size.
This is expected
to obey the scaling form $W\simeq t^{\beta}Z(t/L^z)$;  the
value of $z$ is expected to be 2 as the movement of
masses is diffusive.
Figure 5 shows the scaling plot of $W/t^{\beta}$ versus $t/L^z$ for
four different system sizes $L=16$, $32$, $64$ and $128$ at the
critical point $p\approx 2.35$ for fixed $q=1$. We fix
$z=2$
and find the best collapse of data for $\beta\simeq 0.358$. These
exponent
estimates in $d=1$ are consistent with the scaling relations mentioned
in (a)-(c).

\section{Summary and discussion}
In this paper we have studied a simple lattice model where masses
diffuse and aggregate with rate $1$, unit masses adsorb at any
lattice site with rate $q$ and unit masses desorb from a site
(provided the site is occupied by a mass) with rate $p$. For
$p=0$ (without the desorption process), our model reduces to the
well studied Takayasu model where the steady-state single site mass
distribution has a power law decay, $P(m)\sim m^{-\tau}$ for large $m$
for any nonzero $q$. We show that varying the desorption rate $p$
induces a nonequilibrium phase transition at a critical value $p=p_c(q)$.
For $p<p_c(q)$, $P(m)\sim m^{-\tau}$ for large $m$ as in the Takayasu
($p=0$) case. For $p=p_c(q)$, $P(m)\sim m^{-\tau_c}$ where $\tau_c$ is
a new exponent and $P(m)\sim \exp(-m/m^*)$ for $p>p_c(q)$.
We have solved the model analytically within the mean field theory and
calculated all the mean field exponents exactly. In one
dimension, we have computed the exponents numerically. We have also
presented a general scaling theory.

Our model generalizes the Takayasu model and exhibits
a nontrivial phase transition. There was an earlier
generalization\cite{T} 
of the model where instead of carrying positive masses, the diffusing
particles carried charges $Q$ of either sign while a random charge
$I$, drawn from an
arbitrary distribution, was added with rate $q$
to a lattice site. In this ``charge'' model, the steady-state single site    
charge distribution $P(Q)$ was found\cite{T} to have a power law tail (as
in the mass case), $P(Q)\sim Q^{-\tau}$ for large positive $Q$ when the
mean charge injected was positive, $\langle I\rangle >0$ whereas for
$\langle I\rangle =0$, $P(Q)\sim Q^{-\tau_1}$ for large positive $Q$.
It was shown that the exponent $\tau_1=5/3$ in $d=1$ and $\tau_1=2$
within mean field
theory\cite{T}. Though this change of exponent at a critical value
$\langle I\rangle =0$ is similar to that in our model qualitatively,
the exponent $\tau_c$ of the In-out model is very different from that
of the ``charge'' model. This difference can be traced back to the mass
positivity constraint in the In-out model, i.e., the desorption of
an unit mass can take place
from a lattice site
only if the site has a nonzero mass.

In the In-out model, the total mass is not conserved due to the moves   
involving adsorption and desorption of unit mass. It is interesting to
ask what would happen if the desorption of a unit mass from a site were
followed by adsorption at a neighbouring site, so that the total mass
would be conserved in every move. This was investigated using a lattice
model \cite{MKB} and earlier, within a rate
equation approach \cite{KR}. In this conserved-mass model too there
is a phase transition, but of a different character. It was found that
there is an exponential phase (at high desorption-adsorption rate),
separated by a critical line from a phase with a power-law mass
distribution $P(m) \sim m^{-\tau_{\rm conserved}}$. This distribution
coexists with an infinite aggregate which accommodates a finite fraction
of the total mass --- a real space analog of Bose-Einstein condensation
\cite{MKB}. The exponent $\tau_{\rm conserved}$ was found to be 5/2
within mean field theory\cite{MKB,KR}and $\simeq 2.33$ in 1-d \cite{MKB},
and the same exponent was found to describe $P(m)$ at the critical point.
Evidently, the lack of mass conservation in the In-out model is
responsible for the absence of the infinite aggregate
in its high $q$ phase, as well as the change in the power to
$\tau$ in the Takayasu phase and $\tau_c$ at criticality. 

Another interesting difference between the conserved and the
In-out model is the effect of a preferred direction for the motion of
masses (a mass at site $i$ hops with a higher probability to $i-1$
than $i+1$). We have checked that such a bias does not change the
critical exponents of the In-out model.
However, for the conserved mass model, the bias in
direction changes the value of the exponents at the
transition and in the aggregate phase \cite{MKB}.

The phase transition in the In-out model has some
interesting implications for nonequilibrium wetting transitions 
if we interpret the configuration of masses as an interface profile
regarding $m_i$ as a local height variable. Although the dynamics of the
mass profile in our
In-out model is not physical when interpreted as interface dynamics,
nevertheless the phase transition in our model can be qualitatively
interpreted as a nonequilibrium wetting transition of the interface.
In the In-out model, the fact that the mass at each site is necessarily
non-negative translates into the restriction that the mass profile is
always above a wall at a fixed height (in our case $0$). The presence of
this constraint is the key factor for the wetting transition. At fixed
$p$, as we increase $q$, the mean height $\langle m\rangle$ does not grow
with time as long as $q<q_c$. This is our `Exponential' phase where the
mass or the interface profile is bound to the substrate at zero height.
This phase is also `smooth' as the mean square height fluctuation does not
grow with system size. For $q>q_c$, the interface unbinds from the
substrate and the
mean height $\langle m\rangle\sim vt$ grows linearly with time with a 
velocity $v$. This phenomenon is similar to `wetting' or `depinning' of
interfaces in general. In this `wet' phase (Takayasu phase in our model),
the interface is rough. 
Unlike recently studied models of nonequilibrium wetting, where the
interface in the growing phase is self-affine\cite{hinrichsen,mukamel}, 
our model describes a much rougher
interface for $q>q_c$. At the transition $q=q_c$, though, the interface
is self-affine with a
roughness exponent $\chi= z\beta \simeq 0.7$.

There are various open questions that remain to be settled. In this paper,
we have only studied the phase transition in the steady-state single site
mass distribution function. It would be very interesting to study the
spatial correlations between masses at different sites and to track the
behaviour of mass-mass correlation function as one crosses
the phase boundary in the $p-q$ plane. 

Also in this paper we have only studied the simplest model where
the rates of adsorption, desorption and hopping are constants and
independent of particle mass. An important question
is whether this phase transition would persist for general mass-dependent
rates. In earlier work \cite{Hendricks}, a model with
aggregation, adsorption and desorption was studied, but no
transition to a power-law
phase was found; the difference is traceable to the fact that in that
model, the rate of removal of mass is proportional to the mass, unlike
the unit-mass desorption process considered in the In-out model.
It is therefore highly desirable to identify the class of models with
mass-dependent rates where the phase transition described here will
persist.



\newpage
\begin{figure}
\begin{center}
\leavevmode   
\psfig{figure=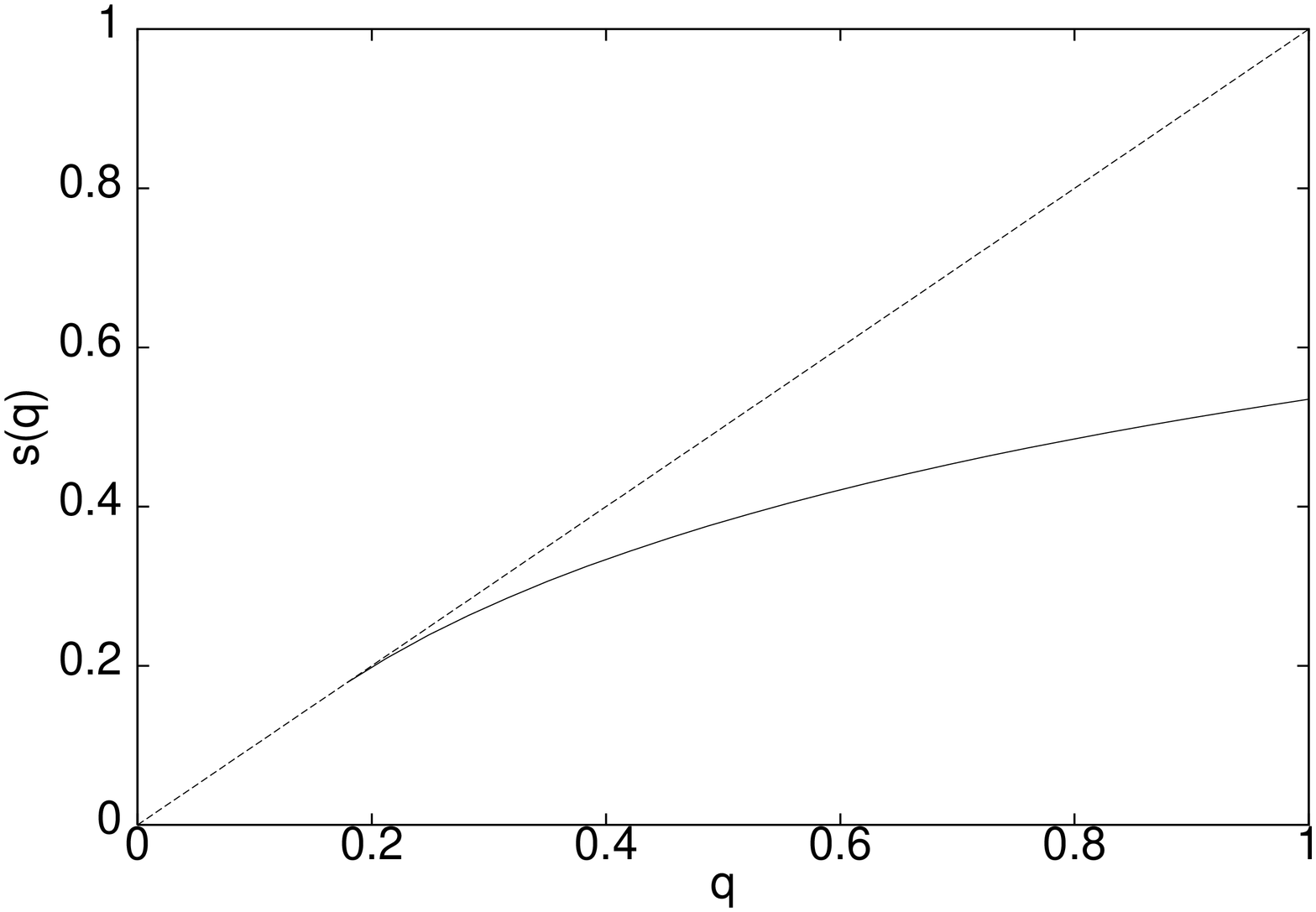,width=12cm,angle=-90}
\caption{The function $s(q)$ as a function of $q$ for $p=1$ (shown  
by the solid line) within the mean field theory. It deviates from the
dotted line ($s(q)=q$) for $q>q_c=3-2\sqrt {2}$.}
\label{fig:sq}
\end{center}
\end{figure}

\begin{figure}
\begin{center}
\leavevmode
\psfig{figure=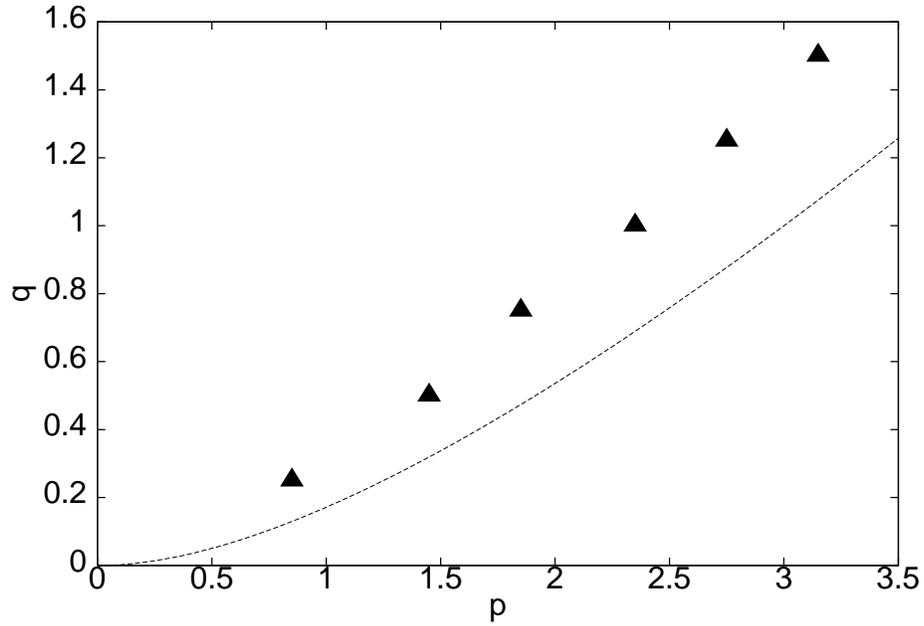,width=12cm,angle=-90}
\caption{The phase diagram of the In-out model in the $p-q$
plane. The dotted line denotes the mean field phase boundary
$q_c(p)=p+2-2\sqrt {p+1}$ and the triangles mark the numerically
obtained critical points in $1$-d.}
\label{fig:phasepq}            
\end{center}
\end{figure}  

\begin{figure}
\begin{center}
\leavevmode
\psfig{figure=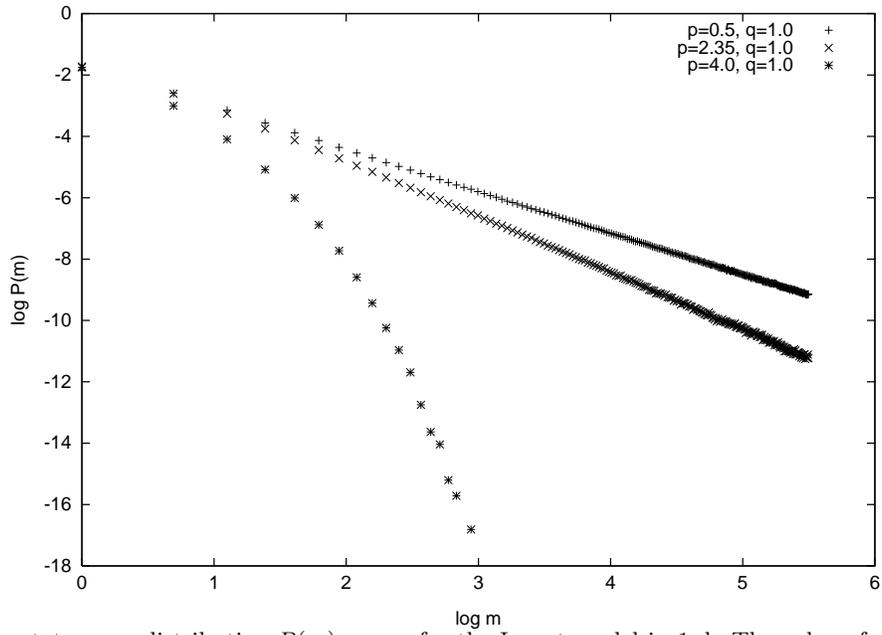,width=12cm,angle=-90}
\caption{The steady-state mass distribution $P(m)$ vs. $m$ for the
In-out model in 1-d. The value of $q$ is kept fixed at $q=1$
and the data is shown for three representative valuesof $p$,
respectively $p <, =, > p_c\simeq 2.35$.}
\label{fig:dist}
\end{center}
\end{figure}  

\begin{figure}
\begin{center}
\leavevmode
\psfig{figure=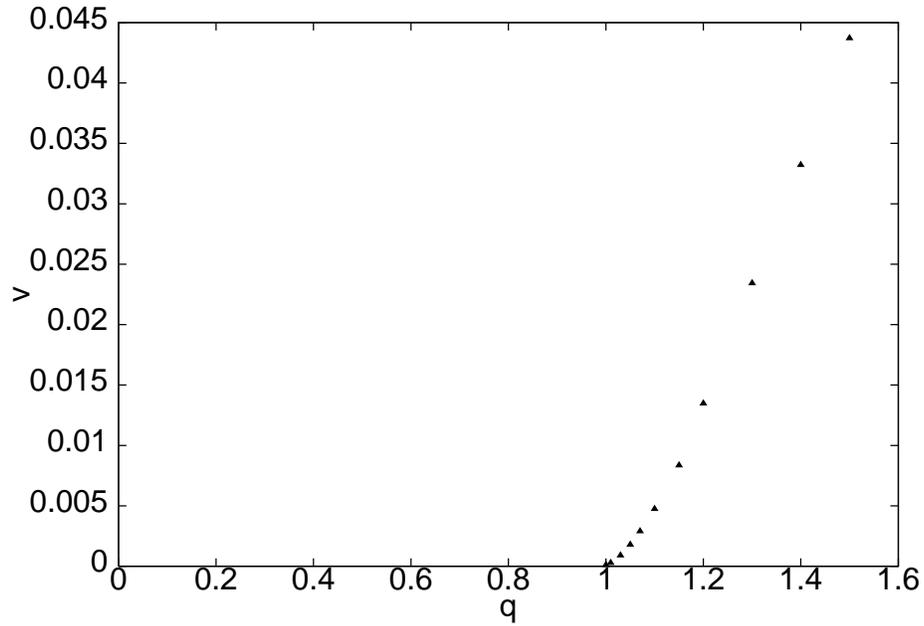,width=12cm,angle=-90}
\caption{ The velocity $v$ as a function of $q$ for fixed $p=2.35$
in $1$-d. The velocity is zero for $q<q_c\simeq 1.0$ and increases
as $(q-q_c)^y$ for $q>q_c$ with $y\simeq 1.47$ in $d=1$.}
\label{fig:vel}
\end{center}
\end{figure}

\begin{figure}
\begin{center}
\leavevmode
\psfig{figure=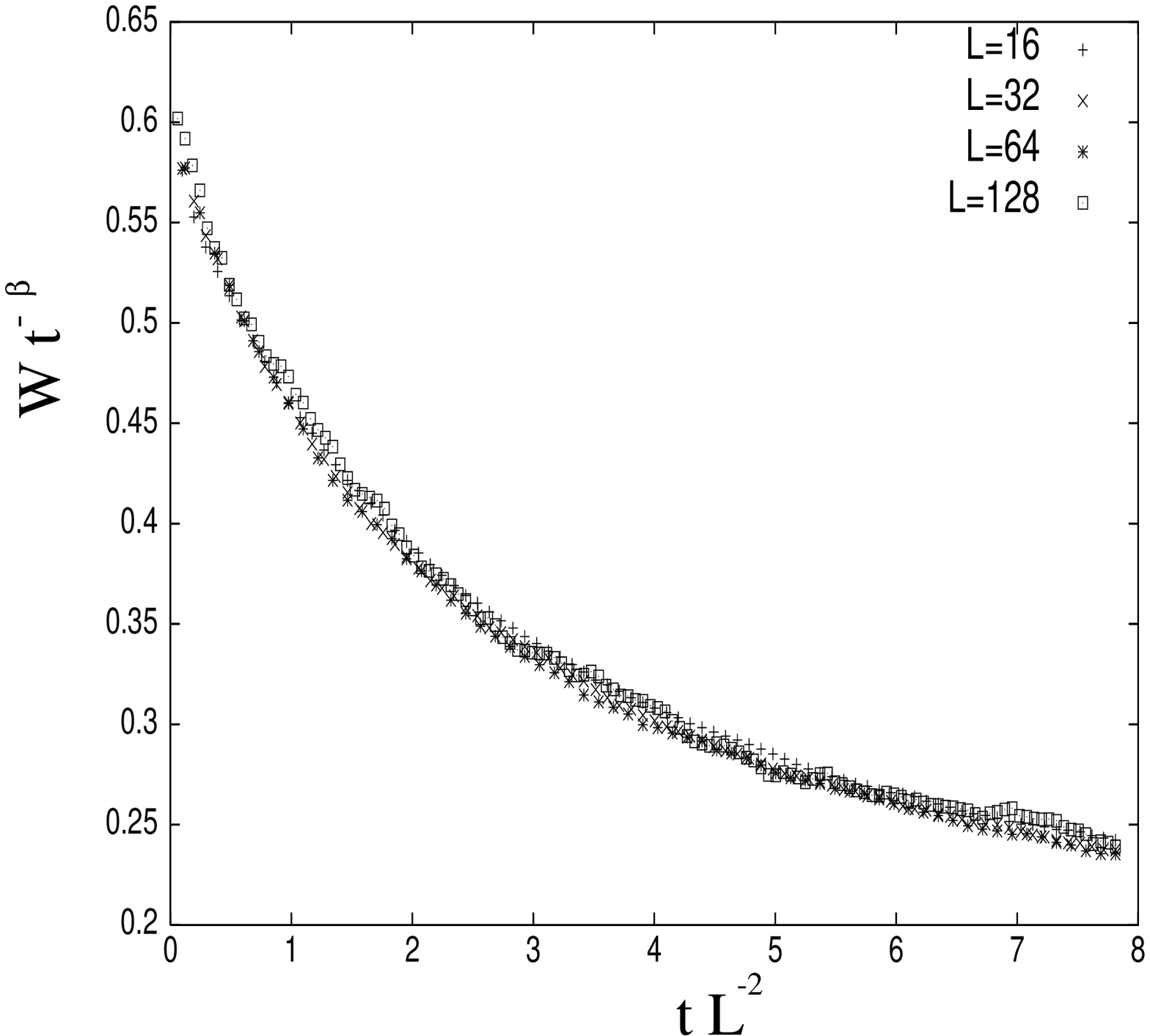,width=12cm,angle=0}
\caption{ Scaling plots of the finite-size studies of width
versus time for four different system sizes $L=16$, $L=32$, $L=64$
and $L=128$ at the critical point $p_c\approx 2.352$ 
with fixed $q=1$. The width is expected to follow the scaling
form $W\approx t^\beta Z(t/L^z)$. The best collapse is obtained
for $\beta\simeq 0.358$ with $z=2$.}
\label{fig:scaling}
\end{center}
\end{figure}



\begin{thebibliography}{999}

\bibitem{BTW} P. Bak, C. Tang and K. Wiesenfeld, Phys. Rev. Lett. {\bf
59}, 381 (1987).

\bibitem{Takayasu} H. Takayasu, I. Nishikawa and H. Tasaki, Phys. Rev.
A {\bf 37}, 3110 (1988); M. Takayasu and H. Takayasu in
{\em Nonequilibrium Statistical Mechanics in One Dimension} ed. V.
Privman (Cambridge Univ. Press, Cambridge, 1997). 

\bibitem{dd} D. Dhar, cond-mat/9909009.

\bibitem{scheidegger} A. E. Scheidegger, Int.
Assoc. Sci. Hydrol. Bull. {\bf 12}, 15 (1967).

\bibitem{liggett} T.M. Liggett, {\em Interacting Particle Systems}
(Springer-Verlag, New York, 1985); R. Durrett, {\em Lecture Notes on
Particle Systems and Percolation} (Wadsworth, Belmont, 1988).

\bibitem{ddrr} D. Dhar and R. Ramaswamy, Phys. Rev. Lett. {\bf 63},
1659 (1989). 

\bibitem{SC} S.N. Coppersmith, C.-h. Liu, S.N. Majumdar, O. Narayan and
T.A. Witten, Phys. Rev. E {\bf 53}, 4673 (1996).

\bibitem{AM} A. Maritan, A. Rinaldo, R. Rigon, A. Giacometti and I. R.
Iturbe, Phys. Rev. E {\bf 53}, 1510 (1996); M. Cieplak, A. Giacometti, A.
Maritan, A. Rinaldo, I.R. Iturbe and J.R. Banavar, J. Stat. Phys. {\bf
91}, 1 (1998).

\bibitem{Melzak} Z.A. Melzak, {\it Mathematical Ideas, Modeling and   
Applications, Vol II of Companion to Concrete Mathematics} (Wiley, New   
York, 1976), p.271.

\bibitem{FF} P.A. Ferrari and L.R.G. Fontes, El. J. Prob. {\bf 3}, Paper
no. 6 (1998).

\bibitem{IKR} S. Ispolatov, P.L. Krapivsky and S. Redner, Eur. Phys. J.
{\bf B2}, 267 (1998).

\bibitem{aldous} D. Aldous and P. Diaconis, Probablity Theory and Related
Fields, {\bf 103}, 199 (1995).

\bibitem{Haff} P.K. Haff, J. Fluid Mech. {\bf 134}, 401 (1983); S.
McNamara and W.R. Young, Phys. Fluids A {\bf 4}, 496 (1992); B. Bernu and
R. Mazighi, J. Phys. A {\bf 23}, 5745 (1990); E. Ben-Naim and P.
Krapivsky, cond-mat/9909176.

\bibitem{krug} J. Krug and J. Garcia, cond-mat/9909034, to appear
in J. Stat. Phys.

\bibitem{RM} R. Rajesh and S.N. Majumdar, cond-mat/9910206, to appear
in J. Stat. Phys.

\bibitem{T4} H. Takayasu, M. Takayasu, A. Provata and G. Huber, J. Stat.
Phys. {\bf 65}, 725 (1991).

\bibitem{MS} S.N. Majumdar and C. Sire, Phys. Rev. Lett. {\bf 71}, 3729
(1993). 

\bibitem{RD} R. Dickman in {\em Nonequilibrium Statistical Mechanics
in One Dimension} ed. V. Privman (Cambridge Univ. Press, Cambridge,
1997).

\bibitem{JK} J. Krug, Adv. Phys. {\bf 46}, 139 (1997).

\bibitem{SZ} B. Schmittmann and R.K.P. Zia, {\it Statistical Mechanics
of Driven Diffusive Systems} Academic, London (1995).

\bibitem{hinrichsen} H. Hinrichsen, R. Livi, D. Mukamel and A. Politi,
Phys. Rev. Lett. {\bf 79}, 2710 (1997).

\bibitem{DEHP} B. Derrida, M.R. Evans, V. Hakim and V.J. Pasquier, J.
Phys. {\bf A26}, 1493 (1993); Physica {\bf A200}, 25 (1993). 

\bibitem{MKB} S. N. Majumdar, S. Krishnamurthy and M. Barma,
Phys. Rev. Lett {\bf 81}, 3691 (1998); see also cond-mat/9908443 (to
appear in J. Stat. Phys.).

\bibitem{KR} P.L. Krapivsky and S. Redner, Phys. Rev. E {\bf 54}, 3553
(1996).

\bibitem{White} W.H. White, J. Colloid Interface Sci. {\bf 87}, 204
(1982). 

\bibitem{gel} R.M. Ziff, J. Stat. Phys. {\bf 23}, 241 (1980).


\bibitem{thinfilms}  B. Lewis and J.C. Anderson, {\it Nucleation and
Growth of Thin Films} Academic, New York (1978). 


\bibitem{T} H. Takayasu, Phys. Rev. Lett. {\bf 63}, 2563 (1989)

\bibitem{mukamel} U. Alon, M. R. Evans, H. Hinrichson and D. Mukamel,  
Phys. Rev. Lett. {\bf 76}, 2746 (1996); Phys. Rev. E {\bf 57}, 4997
(1998).

\bibitem{Hendricks} E.M. Hendricks and R.M. Ziff, J. Coll. Interf. Sc.
{\bf 105}, 247 (1985).

\end{thebibliography}
\end{document}